\documentclass{svjour2}                    % onecolumn
\smartqed  % flush right qed marks, e.g. at end of proof
\usepackage{graphicx}
\usepackage{subfigure}
\usepackage{lscape}
\usepackage{rotating}
\usepackage{color}
 \journalname{Journal of Statistical Physics}
\begin{document}

\title{Hydrophobic and ionic-interactions in bulk and confined water with implications for collapse and folding of proteins
}


\author{S. Vaitheeswaran         \and
        Jie Chen                 \and
        D. Thirumalai
}


\institute{S. Vaitheeswaran \at
           Institute for Physical Science and Technology, University of Maryland, College Park, MD 20742. \\
%              Tel.: +123-45-678910\\
%              Fax: +123-45-678910\\
%              \email{fauthor@example.com}           %  \\
%%%             \emph{Present address:} of F. Author  %  if needed
           \and
           Jie Chen  \at
           Institute for Physical Science and Technology, University of Maryland, College Park, MD 20742. \\
           \and
           D. Thirumalai  \at
           Institute for Physical Science and Technology and\\
           Department of Chemistry and Biochemistry\\
           University of Maryland, College Park, MD 20742.
}

\date{Received: date / Accepted: date}

\maketitle

\begin{abstract}
Water and water-mediated interactions determine thermodynamic and kinetics of protein folding,  protein aggregation and self-assembly in confined spaces. To obtain insights into the role of water in the context of folding problems, we describe computer simulations of a few related model systems.  The dynamics of collapse of eicosane shows that upon expulsion of water the linear hydrocarbon chain adopts an ordered helical hairpin structure with 1.5 turns. The structure of dimer of eicosane molecules has two well ordered helical hairpins that are stacked perpendicular to each other. As a prelude to studying folding in confined spaces we used simulations to understand changes in hydrophobic and ionic interactions in nano droplets. Solvation of hydrophobic and charged species change drastically in nano water droplets.  Hydrophobic species are localized at the boundary. The tendency of ions to be at the boundary where water density is low increases as the charge density decreases. Interaction between hydrophobic, polar, and charged residue are also profoundly altered in confined spaces.  Using the results of  computer simulations and accounting for loss of chain entropy upon confinement we argue and then demonstrate, using simulations in explicit water, that  ordered states of generic amphiphilic peptide sequences should be stabilized in cylindrical nanopores.
\end{abstract}

%%%%   TEMPLATE   %%%%
%\section{Introduction}
%\label{intro}
%Your text comes here. Separate text sections with
%\section{Section title}
%\label{sec:1}
%and \cite{RefJ}
%\subsection{Subsection title}
%\label{sec:2}
%as required. Don't forget to give each section
%and subsection a unique label (see Sect.~\ref{sec:1}).
%\paragraph{Paragraph headings} Use paragraph headings as needed.
%\begin{equation}
%\end{equation}
%
%% For one-column wide figures use
%\begin{figure}
%% Use the relevant command to insert your figure file.
%% For example, with the graphicx package use
%  \includegraphics{example.eps}
%\end{figure}
%%
%% For two-column wide figures use
%\begin{figure*}
%% Use the relevant command to insert your figure file.
%% For example, with the graphicx package use
%  \includegraphics[width=0.75\textwidth]{example.eps}
%\end{figure*}
%\begin{table}
%\begin{tabular}{lll}
%\hline\noalign{\smallskip}
%\end{tabular}
%\end{table}
%%%%   TEMPLATE   %%%%

\section{Introduction}
In the process of protein folding an ensemble of unfolded states reach
compact folded structures. Protein folding  has become a paradigm problem in molecular biology, and
has inspired a large number of studies using numerous methods
\cite{Thirumalai10ARB,Onuchic04COSB,Shakhnovich06ChemRev,Dill08ARB,Schuler08COSB,Thirumalai05Biochem}.
The cited reviews have shown that a multi-faceted approach is needed to understand global aspects of protein folding.
Although deciphering the global principles of folding cannot be achieved using brute
computer simulations alone, specially designed computers that are capable of generating long trajectories, are starting to  reveal molecular details of the folding of small single domain proteins and the role of water in well-defined systems \cite{Shaw10Science}. Historically, considerable understanding of  the nature of driving force for protein self-assembly have been obtained simple model systems.
In this review, we describe computer simulation results for simple systems, which illustrate the role water
in  protein collapse and folding in confined spaces.

Typically hydrophobic residues are buried in the interior of folded proteins, and hence are shielded from water.  The predominant driving force in protein folding is  hydrophobic  interaction between non-polar residues \cite{Tanford_AdvProtChem70,Dill_Biochem90,Baldwin_JMB07}, which is reasonable because globular proteins  contain $\sim$ 55\%  hydrophobic residues \cite{Camacho93PRL}.  Thus, understanding the hydrophobic effect, and more generally water-mediated interactions, in the context of protein folding and aggregation \cite{Straub10COSB,Straub11ARPC,Shea11COSB} holds the key in describing their self-assembly.   Hydrophobic interaction refers to the free
energy increase upon transfer of non-polar species from non-polar
solvents to water. The reluctance of water to allow rupture of  the persistent but dynamically changing complex network prevents non-polar species from being easily solvated. To dissolve small  ($<$ 1 nm) non-polar solutes, a cavity of appropriate size and shape needs to be created in water, which is free energetically unfavorable.
 Water-mediated interactions between solute molecules play a crucial role in
protein folding \cite{Dill_Biochem90,Liu_Berne_Nat05}.
Interactions between small hydrophobic solutes like methane have served as
useful stand-ins for those between amino acid side chains in aqueous solution
\cite{Cheung_Onuchic_PNAS02}.
Such interactions, quantified using potentials of mean force (PMFs) averaged over
solvent configurations, have been extensively studied in bulk water
\cite{Pangali_Berne_JCP79,Chandler_Nat05}.
However,  globular proteins are polymeric and undergo collapse transition from unfolded states, and hence polymer chains that can adopt globular conformations in water are good model systems for probing the process of collapse.
The first part of the perspective deals with new results for collapse of a single hydrocarbon chain and the association of two chains.
These studies are models for describing the early stages of protein collapse and aggregation driven  by hydrophobic forces as is the case for A$\beta$-peptides.

In the second part, we turn our attention to changes in molecular
interactions between amino acid side chains in confined spaces.
Besides their relevance in materials science, confinement effects play a major role
in protein folding \textit{in vivo}, e.g., in chaperonin-assisted folding
\cite{Thirumalai_Lorimer_ARBBS01}, co-translational folding in the exit tunnel
of the ribosome \cite{Woolhead_Johnson_Cell04,Ziv_Thirumalai_PNAS05}, and dynamics
in the crowded cellular environment \cite{Minton_COSB00,Cheung05PNAS,Kudlay09PRL}.
Several experimental
\cite{Eggers_Valentine_JMB01,Eggers_Valentine_ProtSci01,Ravindra_JACS04,Campanini_Bettati_ProtSci05,Bolis_JMB04},
theoretical and computational studies
\cite{Betancourt_Thirumalai_JMB99,Zhou_Dill_Biochem01,Klimov_Thirumalai_PNAS02,Ziv_Thirumalai_PNAS05,Cheung_Thirumalai_JMB06,Lucent_Pande_PNAS07,Zhou_JCP07}
have examined changes in the confinement-induced stability of proteins.
General theoretical considerations and explicit simulations using coarse-grained
models for polypeptide chains have shown that the subtle balance of many factors,
including hydrophobic and ionic interactions in confined water, entropic
restrictions of the conformations of polypeptide chains, and specific interactions
between amino acid residues and the confining boundaries determine the stability of
confined proteins \cite{Betancourt_Thirumalai_JMB99,Cheung_Thirumalai_JMB06,OBrien08NanoLett}.
In many cases,  stabilization of the folded state of proteins in confined spaces
relative to bulk solvent \cite{Eggers_Valentine_JMB01,Ravindra_JACS04,Campanini_Bettati_ProtSci05,Bolis_JMB04} can be explained by the entropic
stabilization mechanism (ESM).  According to ESM  reduction in the allowed
conformations of the unfolded states of proteins in confined  leads to stabilization of the native state with respect to the bulk
\cite{Betancourt_Thirumalai_JMB99,Zhou_Dill_Biochem01,Klimov_Thirumalai_PNAS02,Ziv_Thirumalai_PNAS05}.
However, confinement can also destabilize the folded state
\cite{Cheung_Thirumalai_JMB06,Lucent_Pande_PNAS07} due to the alterations in
hydrophobic interactions, which can result in a net attraction between the protein
and the confining boundary \cite{Eggers_Valentine_JMB01,Eggers_Valentine_ProtSci01}.
To understand confinement effects systematically we describe the changes in
hydrophobic and ionic interactions using model systems and side chains (SCs) of
amino acid residues in spherical and cylindrical cavities, and expound on the
implications of these findings for folding in confined spaces.

\section{Methods}

\subsection{Hydrocarbon chains in water}
{\bf Models:} We give a brief description of the simulations reported here for the dynamics of collapse and association of hydrocarbon chains. Intermolecular interactions are modeled using standard
site-site potentials. The pair potentials include partial charges and
Lennard-Jones (LJ) interactions. The charges are located on the centers
of the atoms. Site-site interaction between molecules i and j, with R
denoting collectively the coordinates of the molecule, is given by

\begin{equation}
V(R_i,R_j)=\sum^{atoms}_{\alpha\in i,\beta\in j}{\frac{C_{\alpha\beta}^{12}}{r_{\alpha\beta}^{12}}-\frac{C_{\alpha\beta}^6}{r_{\alpha\beta}^6}+\frac{q_{\alpha} q_{\beta}}{4\pi\epsilon_0r_{\alpha\beta}}}
\end{equation}
where $r_{\alpha\beta}$ is the scalar distance between sites $\alpha$ and $\beta$. With this standard
parametrization the interaction between molecules is fully specified in
terms of the LJ potentials and the partial charges $q_\alpha$.

{\bf Water:} We use the three-site rigid SPC/E model for water in which the hydrogen atoms are at a distance
0.1 nm from the oxygen site and the HOH angle is 109.47$^{\circ}$. The charges
on oxygen and hydrogen are $q_O= - 0.8476e$ and $q_H= 0.4238e$, respectively.
The LJ parameters for the oxygen sites are $\epsilon_O=0.1548$ kcal/mol and $\sigma_O=0.3166$ nm.

{\bf Hydrocarbon (HC):} Following our previous studies, we used a united
atom, site-site interaction model for the linear alkane chain $CH_3-(CH_2)_n-
CH_3$ with $n=18$. The intramolecular force field for the hydrocarbon
chain can be decomposed into $V_{intra}=  V_2 + V_3 + V_4 + V_5$. The first
two terms correspond to harmonic bond stretching and harmonic bond
angle potential. The term $V_4$ contains the torsion interactions involving
four consecutive sites. The $V_5$ term represents non-bonded interactions
between sites that are separated by three or more sites along the chain. The non-bonded
terms are modeled using Lennard-Jones potentials. The parameters for
the HC force field are given elsewhere \cite{SmitJCP1995}.

{\bf Simulation Details:} The system consisted of 1995 (1942) solvent
molecules plus 1 alkane chain molecule (2 alkane chain molecules).
The equations of motion were integrated using an iterated form of the Beeman algorithm
with a time step of 1 fs.
Periodic boundary conditions were imposed in all three spatial dimensions.
The volume of the system was set so that the water density was 1 g/cm$^3$, and the
temperature of the system was maintained at 25 C by Nose-Hoover thermostat.
At the start of the simulations, the alkane chain was placed in contact
with the solvent and the solvent molecules were allowed to adapt to
the presence of the alkane before any motion of the alkane was permitted.
In this way, the disruption of the solvent due to the presence
of the solute was allowed to relax.
Once the systems were stabilized, production runs were made for long enough
duration so that both collapse of a single chain and association of two chains
could be observed.

\subsection{Small molecules and amino acid side chains in water}
We used Metropolis Monte Carlo \cite{Metropolis_JCP53} simulations to study the
energetics of aqueous solvation of methane molecules and the model ions
M$_{q+}$ and M$_{q-}$ in spherical and cylindrical water--filled nanopores using the TIP3P model for water \cite{Jorgensen_Klein_JCP83} and a unified atom
representation for methane \cite{Kalra_Hummer_Garde_JPCb04}.
Model ions M$_{q\pm}$ were created by adding charges of magnitude
$q^+ = |q^-| = 0.4e$ or $e$ to the methane spheres, where $e$ is the electronic
charge \cite{Wallqvist_Thirumalai_JACS98}.
In order to assess the importance of charge density ($\zeta$) in affecting solvation we
also studied, for the 3 nm spherical droplet, ions M*$_{q+}$ and M*$_{q-}$ with
$q^+ = |q^-| = e$ and twice the volume of the methane sphere.

%\textbf{Simulation Details:}
Simulations were performed at constant number of molecules $N$, total volume
$V$ and temperature $T=298$ K, and thus sampled the canonical ensemble.
The effective volume available to the water molecules is taken to be
\begin{equation}
  V_{\mathrm{eff}} = V - N_sV_s
\end{equation}
where $N_s$ is the number of solute molecules in the pore, and $V_s$ is the
excluded volume due to each solute.
The spherical droplets had diameters ($D$) in the range 1.0--4.0 nm.
The cylindrical pores had diameters ($D$) in the range 0.8--20 nm, and lengths
$L$ varying from 1.2 to 8.3 nm.
These volumes were bounded by hard walls, with the potential energy at the
wall being large enough ($10^{12}$ kJ/mol) to confine the system to the
desired volume.
To simulate confinement effects, no periodic boundary conditions are applied.

In the spherical droplets, the free energy of a single solute molecule was
calculated as a function of its radial distance from the center of the droplet
using Monte Carlo simulations with umbrella sampling and the weighted histogram analysis method (WHAM)
\cite{Kumar_JCompChem92}.
Umbrella sampling and WHAM were also used to calculate the PMFs between pairs
of solutes in the spherical droplets.
In the cylindrical pores, the same technique was used to calculate interaction
free energies, $-k_BT \log P(r)$, where $k_B$ is Boltzmann's constant and
$P(r)$ is the probability of finding the two solutes a distance $r$ apart.
Because of the quasi one-dimensional nature of the confinement when $r \gg D$,
we do not subtract the free energy contribution $-2k_BT \log r$ which arises
from the increase in phase space proportional to $r^2$ in spherically symmetric
systems.
Therefore, these profiles cannot be compared directly to PMFs in bulk or in the
spherical droplets.

\textbf{Molecular dynamics simulations:}
The interaction between the side chain pairs ALA-PHE, SER-ASN and LYS-GLU was
studied \cite{Vaitheeswaran_PNAS08} using molecular dynamics simulations with
the CHARMM22 force field \cite{Brooks_Karplus_JCC83} and TIP3P water
\cite{Jorgensen_Klein_JCP83}.
In each amino acid, only the side chain was retained.
The carboxyl and amino groups were deleted and the C$_{\alpha}$ atom was
replaced by a hydrogen atom.
Thus, for instance, the side chain of ALA was represented by methane and PHE by
toluene.

In the cylindrical pores, interaction free energies, $-k_BT \log P(r)$
($k_B$ is Boltzmann's constant, $r$ is the distance between the centers of mass
of the SC pair, and $P(r)$ is the probability of finding the two solutes at a
separation $r$), were calculated at a fixed volume using using the adaptive biasing force (ABF) technique \cite{Darve_Pohorille_JCP01,Henin_Chipot_JCP04} implemented in NAMD \cite{Phillips_Schulten_JCC05}.
Because of the quasi one-dimensional nature of the confinement when $r \gg D$,
we do not subtract the free energy contribution $-2k_BT \log r$ which arises
from the increase in phase space proportional to $r^2$ in spherically symmetric
systems.
Therefore, these profiles cannot be directly compared to PMFs in bulk.

\section{Results and Discussion}

\subsection{Hydrocarbon Chains in Water:}

{\bf Hydration of Octane:}
The probability of creating a large anisotropic cavity to accommodate a solute would decrease with the volume of the solute. Thus, one might expect that large solutes would distort the hydration structure at least locally. Such an expectation is borne out in the studies of water near a hydrophobic wall, which showed that close to the wall there is a loss of one hydrogen bond per water molecule \cite{Lee84JCP,Lee94JCP}. A natural question is how the solvent responds to the presence of large apolar molecules with internal degrees of freedom like linear hydrocarbon chains  $CH_3-(CH_2)_{n}-CH_3$, which are interesting model systems for studying the early stages of  collapse of proteins.
As a result, a number of studies focusing on various aspects of hydrocarbon behavior
in water have been reported \cite{MountainJACS2003,Chakrabarty09JPCB,Ferguson10PNAS}.
We first discuss the case of octane relegating the discussion of hydrocarbon collapse to the next section.

Because linear hydrocarbon molecules are intrinsically flexible and can undergo rotational motion around the dihedral angles populating \textit{cis}, \textit{trans}, and \textit{gauche} states,  one may expect that a solute with the size of octane ($n$ = 6) or dodecane ($n$ = 10) could locally distort the structure of water.  Indeed, if octane or dodecane were treated as a sphere the probability of creating a void large enough enough to accommodate them is exceedingly low. Molecular simulations studies of solubility of large hydrocarbon molecules revealed several interesting findings.
(1) Using the SPC/E model for water and a coarse-grained hydrocarbon model described
in \cite{SmitJCP1995}, Mountain and Thirumalai \cite{Mountain98PNAS} showed that
there are minimal changes in the tetrahedral network of water structure even near the
hydration layer of octane.
Although changes  in water structure was not reported in simulations of dodecane
\cite{Wallqvist} it was found that the extended structure was favored implying that
the network of hydrogen bonds is sufficiently plastic to accommodate a large solute as large as dodecane.
(2) Comparison of the distributions, $P(cos \theta)$, ($\theta$ is the angle between three near neighbor oxygen atoms) between water in the bulk and oxygen atoms within 0.5 nm of at least one of the carbon atoms in octane shows peaks around the tetrahedral value ($cos \theta \approx \frac{1}{3}$) \cite{Mountain98PNAS}, which shows that a cavity of an appropriate shape can be found to accommodate octane.
The pair functions $g_{OH}(r)$ for bulk water and those around the hydration shell
of water show that positions and widths of the two peaks are similar (see Fig.  4 in \cite{Mountain98PNAS}.
(3) The shape of the cavities in which the solute is housed are best described as oblate ellipsoids.  Using $\lambda = \frac{R_3}{R_1}$, where $R_3$ and $R_1$ are the major and minor radii of gyration of the solute, as a measure of anisotropy it was found that $\lambda$ varies from 1.0 for methane and ethane to 1.3 for octane \cite{Mountain98PNAS}.
(4) Several lines of evidence showed that octane and dodecane adopt extended structures in water. For collapse to occur a substantial fraction of dihedral angles must be in the \textit{gauche} conformation. However, the probability of finding conformations with appreciable number of \textit{gauche} conformations is small. In addition, the mean end-to-end distance is similar to that found in the gas phase, which implies that water does not force collapse of hydrocarbon chains even with $n=$ 10.

{\bf Collapse and association of Hydrocarbon chains in water}
A Flory-like theory showed that when $n$ is $\sim$ 20 a single HC would collapse and adopt compact structures \cite{MountainJACS2003}. The estimate was made using average interaction per contact between the interaction centers, and hence is likely to be only approximately correct.
Explicit simulations, first carried out by Mountain and Thirumalai
\cite{MountainJACS2003}, confirmed that collapsed states can be realized with $n$ = 18.
However, subsequent studies \cite{Athawale07PNAS,Chakrabarty09JPCB,Ferguson10PNAS,Ferguson09JPCB} showed that, depending on the force field, realization of stable collapsed structures might require $n \ge$ 20. Somewhat surprisingly, several studies have shown the HC in the collapsed structures are not merely maximally compact but adopt regular structures ranging from helix hairpin (\textbf{HH}), toroids and other complex folds  depending on $n$ \cite{MountainJACS2003,Chakrabarty09JPCB,Ferguson10PNAS}. Thus, linear hydrocarbon chains, which typically become compact on nanosecond time scales, are ideal model systems for probing some aspects of protein collapse.

We first  report preliminary   molecular dynamics simulations of the dynamics of collapse of  a single hydrocarbon chain (HC) $CH_3-(CH_2)_{18}-CH_3$ in water using the models described in the Methods section.
To illustrate the pathways explored by the HC we show in
Fig.~\ref{fig:snapshots}a the structures
explored in the transition from extended structures to a compact conformation in a
trajectory with first passage time on the order of 0.25 ns.
Although,  Fig.~\ref{fig:snapshots}a  shows a sequential
transition the movie (can be viewed at http://www.youtube.com/watch?v=dWW83YVhX-U
) shows that transitions between extended and collapsed structures
occur prior to formation of the compact conformation. It appears that for collapse of extended HCs water expulsion occurs on time scales that are
significantly less than $\tau_c$, the overall collapse time.
The radius of gyration of the final conformation of HC chain is consistent with
Flory's scaling law $\sim N^{\nu}$, where $N$ is number of beads in the HC chain
and $\nu=1/3$, and with a pre-factor that is close to the LJ $\sigma$.
The structure of the compact structure is best characterized as a 1.5 turn \textbf{HH}.

In order to obtain a picture of HC association we  also conducted preliminary
studies of interaction between two identical eicosane HCs.
In the generated trajectory  we observed collapse of each HC chain and subsequent
association (see http://www.youtube.com/watch?v=KP8bP6jWV40 for a movie of the trajectory).
We should emphasize that this is only one of the possible pathways
and more extended simulations are needed to fully determine the mechanism by which
HCs associate.
Similar to the case of single HC chain, the collapsed hairpin conformation can
become extended (see the cyan chain in Fig.~\ref{fig:snapshots}b), and re-collapses
to the more condensed conformation, where the two HC chain tends to stack together
with the long axis of one hairpin being perpendicular to the long axis of the other.
This resulted in a compact structure in which each HC adopts a \textbf{HH} conformation.
It is likely that in this pathway water is expelled early in the association
process, just as in oligomer formation between A$\beta$ fragment with predominantly hydrophobic residues \cite{Klimov04PNAS}, leading to the formation of a ``dry'' hydrophobic interface.

\subsection{Effects of Nanoconfinement:}

\textbf{Water structure in nanopores:}
The water structure in the nanopores is considerably perturbed by the presence
of  nonpolar walls.
Water hydrogen bonds are broken in the layer adjacent to the confining boundary resulting in
water being more loosely bound in the surface layer compared to
the bulk phase.  Water molecules are oriented with the molecular dipoles nearly orthogonal
to the droplet radius leading to  vapor-like, low density layer adjacent to the walls, which as we
describe below, can easily accommodate small hydrophobic species such as methane
\cite{Vaitheeswaran_JACS06}.

\textbf{Methane is localized at the surface of a spherical cavity:}
Disruption of water structure adjacent to the walls makes the surface
region strongly favorable for nonpolar solutes like methane.
Figs.~\ref{fig:free_energy_D1-3} a and b show the free energy of a single methane
molecule (black, filled circles) in spherical droplets of diameters 3 and 4 nm
respectively.
In both droplets the surface is more favorable for the methane than the
interior by $\sim 10-15$ kJ/mol ($4-6$ $k_BT$ at 298 K)
\cite{Matubayasi_Levy_JPC96,Henin_Chipot_JCP04,Ashbaugh_Pethica_Lang03,Vaitheeswaran_JACS06}.
In the largest and the most bulk-like droplet, the free energy difference of
$\sim 10$ kJ/mol is close to the hydration free energy of a methane molecule
calculated from simulations and experiment \cite{Henin_Chipot_JCP04}.
This supports the view of the droplet surface inducing a vapor-like state of
low density with broken hydrogen bonds \cite{Wallqvist_Levy_JPCb01}.
Because of the broken H-bonds at the surface, the solvent loses
less entropy due to ordering of hydrogen bonds around the methane if the
nonpolar solute is confined to the droplet surface.
The physical picture is supported by calculations, which show that the free energy of
solvation of methane in spherical  droplets is strongly dominated by entropic
contributions \cite{Vaitheeswaran_JACS06}. Remarkably, our simulations and theory show that methane would be pinned at the surface regardless of the size of the spherical droplet.

\textbf{Ion solvation depends on sign of charge and size of the spherical droplet:} The solvation free energies of positive
(red, plus signs) and negative (blue, filled triangles) ions, M$_{q\pm}$, that
are constructed by assigning charges of magnitude $0.4e$ to methane molecules are also shown in Figs.~\ref{fig:free_energy_D1-3} a and b.
These fictitious ions have a greatly reduced surface propensity
compared to methane.
Unlike the nonpolar solutes, the free energies of the ions strongly depend on
the droplet size.
In droplets of diameter 3 nm and less, the ions are preferentially found at the
surface.
In the $D=4$ nm droplet (Fig.~\ref{fig:free_energy_D1-3}b) the interior is
nearly bulk-like \cite{Vaitheeswaran_JACS06}, and is more favorable than the
surface for both positive and negative ions.
Thus, the penetration of the ions into the droplet interior increases with
decreasing curvature of the droplet surface, in accord with the findings for the chloride ion \cite{Stuart_Berne_JPCa99}.
Just as in bulk water \cite{Lynden-Bell_Rasaiah_JCP97} there is an asymmetry
between the behavior of cations and anions.
%In addition, the magnitude of the effect depends on whether the ion is
%positively or negatively charged.
Anions have a stronger tendency for surface solvation than cations with the
same charge magnitude.
As $D$ increases, the enthalpy gain due to the solvation of the ions is greater
than the entropy loss due to ordering of water molecules around this solute.
The enthalpy-entropy balance depends on the interplay between the hydrophobicity
and the ionic nature of the solutes.

\textbf{Hydration depends on charge density:} To probe the balance between hydrophobic and electrostatic interactions, we
calculated the free energies of  ions, M$_{q\pm}$, with charge magnitude $e$,
and also the ions M*$_{q\pm}$ which have the same charge magnitude, but with
twice the volume of a methane molecule (Fig.~\ref{fig:free_energy_D1-3}c).
Thus, each starred ion has half the value of v$\zeta$ of the corresponding
unstarred ion.
The ions M$_{q\pm}$ have a strong (enthalpic) preference for the interior
over the surface of the droplet, with the free energy difference being of
the order of $10-15$ $k_BT$.
Molecular dynamics simulations with a non-polarizable force field
\cite{Marrink_Marcelja_Lang01} have obtained similar values for the the free
energies of sodium and chloride ions in water, near hydrophobic or purely
repulsive surfaces, and also at a water liquid-vapor interface.
For M*$_{q+}$ and M*$_{q-}$, the surface is much less unfavorable compared
to M$_{q+}$ and M$_{q-}$ respectively.
The free energy profile for M*$_{q-}$ illustrates the balance between
hydrophobicity and ionic character.
The anion M*$_{q-}$ preferentially resides, not in the interior or at the surface, but
0.2 nm below the surface, due to its competing tendencies for surface and
interior solvation.

The asymmetry in the solvation of cations and anions in shperical water droplets, along
with the $\zeta$ dependence also reflects the differences in the
solvation of sodium and halide ions in water clusters and slabs
\cite{Jungwirth_Tobias_JPCb02}.
Sodium and fluoride ions were found to be solvated in the interior while the
larger halides had a propensity for surface solvation in the order
Cl$^- <$ Br$^- <$ I$^-$  \cite{Jungwirth_Tobias_JPCb02}.
The surface propensities of solutes at an air-water interface depend on the
balance between hydrophobic (excluded volume) and charged interactions.
This balance can be tuned by adjusting the $\zeta$  (or by adjusting
the van der Waals parameters, keeping the charges fixed).
Decreasing $\zeta$  increases the tendency for surface solvation.
For a given $\zeta$, anions have a greater preference for the surface
than cations. Thus, a single parameter ($\zeta$) tidily explains the nature of solvation of spherical ions at interfaces.

\textbf{Water-mediated interactions between solutes in spherical and
cylindrical pores:}
Since  methane molecules are confined to the surface of water droplets,
their mutual solvent mediated interaction should reflect the
disruption in the water structure due to the confining walls. Potentials of mean force (PMFs) between two
methane molecules in droplets at various $D$ values (Fig.~\ref{fig:pmf}a) show an increased
tendency for the methanes to associate compared to bulk water \cite{Shimizu_Chan_JCP00}.
Surprisingly, we find that the secondary solvent separated minimum (SSM)
that appears in the bulk profile is completely absent, even when $D=4$ nm.
The absence of the solvent separated minimum in the $D=4$ nm droplet, which is
entirely due to the presence of boundaries, is intriguing because the
properties of water in large droplets are bulk-like in all crucial respects
\cite{Vaitheeswaran_JACS06}.
The  second minimum in the bulk corresponds to configurations where
the methanes are separated by a single water molecule that is hydrogen bonded
to other waters.
In the immediate vicinity of the confining surface, where
bonds between water molecules are disrupted, such configurations are strongly unfavorable.
The preference for being pinned at the surface is consistent with a view that
a methane is likely to be localized in regions with maximum unsatisfied hydrogen
bonds, which  is readily realized at the boundary.
A striking feature of Fig.~\ref{fig:pmf}a is that the calculated PMFs are
independent of the size of the droplet. This is because methanes approach
each other along the surface which is energetically similar in all the droplets
\cite{Vaitheeswaran_JACS06}.

Fig.~\ref{fig:pmf}b shows the PMFs between the ions M$_{q\pm}$ with a charge
magnitude of $0.4e$.
Reflecting the surface propensities, which depend on the droplet size
(Figs.~\ref{fig:free_energy_D1-3}a and b), the solvent-mediated interaction
between the oppositely charged ions also varies strongly with the droplet size.
As the size of the water droplet increases, the interior becomes more
bulk-like and for the largest droplet, the profile is very similar to that for
two methanes in bulk water, with a solvent separated minimum at a separation of
$\sim0.7$ nm.
In the 2 and 3 nm droplets, the secondary minimum is destabilized by $\sim 2$
kJ/mol relative to the 4 nm droplet, reflecting the greater role of confinement.

The role played by the shape of the confining volume in determining the
solvent-mediated interactions between two methane molecules is
illustrated in Fig.~\ref{fig:cyl_d1.6}, for a cylindrical water-filled pore of
diameter $D=1.6$ and length $L=2.08$ nm.
The calculated interaction free energy (solid, black line) shows the
characteristic contact minimum and two distant minima corresponding to the pore
dimensions \cite{Vaitheeswaran_JCP09}. Parenthetically,
we emphasize that the free energy profile in the cylindrical pore cannot be
directly compared to the PMFs in bulk or in the spherical droplets (see Methods).
The distant minima arise due to the preferential solvation of the methanes at
the surface of the cylindrical pore, just as in the spherical droplets.
Treating the spherically symmetric methane molecules as point objects that are
strictly confined to the pore surface, we also calculated the contribution from
the translational entropy of the hydrophobes, $-T \Delta S^{\mathrm{A}}$.
This purely geometric term is shown by the dotted blue curve in
Fig.~\ref{fig:cyl_d1.6}.
As with the total free energy, $-T \Delta S^{\mathrm{A}}$ is only determined to
within an additive constant, and is vertically aligned with the first distant
minimum in the free energy at $r \approx$ min\{$D,L$\}.
Fig.~\ref{fig:cyl_d1.6} shows that $-T \Delta S^{\mathrm{A}}$ qualitatively
captures the curvature of the free energy profile.
Thus, the solute translational entropy favors the first distant minimum in the
free energy, while the solvent drives the solutes to either minimize or maximize
their separation.
Qualitatively similar free energies of interaction have been calculated for
the nonpolar pair of amino acid side chains ALA-PHE in cylindrical pores
\cite{Vaitheeswaran_PNAS08}.

\textbf{Balance between hydrophobic and electrostatic forces in
LYS-GLU interactions:}
Many eukaryotic proteins are intrinsically
disordered, which implies that they do not adopt well defined structures in
isolation but do so only upon interaction with other partners.
In general such proteins contain charges, and hence understanding water-mediated
interactions between charged species both in bulk and in confined spaces determine  their conformational properties.
The interactions between charged residues (Lys and Glu for example) also involve
a subtle balance between hydrophobic and electrostatic interactions.
The side chains of lysine and glutamate, which carry charges of +e and -e, respectively,
at physiological pH, also have nonpolar parts that prefer to be sequestered from the water.
Therefore, the LYS$^+$-GLU$^-$ pair interaction can be mostly electrostatic or both
electrostatic and hydrophobic, depending on their relative orientations.
The PMFs between LYS and GLU in the $\rightarrow \leftarrow$ and the
$\uparrow \uparrow$ orientations (see Fig.~\ref{fig:Lys-Glu}a and stick figures
in b and c) in the bulk show well defined contact and solvent separated minima.
These configurations are stabilized in contact by $8.5-10$ kJ/mol relative to
extended separations.
In the nanopores, the $\rightarrow \leftarrow$ pair, which is predominantly
electrostatic, has the same contact and solvent separated minima at $\sim 0.6$
and $\sim 0.8$ nm respectively, as in bulk solvent (Fig.~\ref{fig:Lys-Glu}b).
In this orientation, the free energy has another minimum corresponding to the
state where the two molecules are approximately a pore length apart,
with their charged ends hydrated and the nonpolar ends at the surface.
The free energy in the $\uparrow \uparrow$ orientation shows features of both
electrostatic and hydrophobic interactions.
This pair is strongly driven to contact at $\sim 0.43$ nm.

The LYS-GLU system illustrates how confinement alters the balance between
hydrophobic and electrostatic interactions.
In the absence of charges, the interaction between these side chains would be entirely
hydrophobic.
If this were the case, then the free energy of interaction between these
hypothetical molecules would have two distant minima in addition to the contact
minimum, at separations corresponding to the pore dimensions, similar to the
case of the two methanes and the ALA-PHE pair \cite{Vaitheeswaran_PNAS08}.
If this was a purely electrostatic pair, the solutes will be fully hydrated and
the free energy profile will have a contact minimum and an SSM, but no distant
minima.
In this scenario, the presence of the confining boundaries will have only a
minimal effect on the interaction between the solutes.
Because favorable electrostatic interactions are expected to dominate in the
bulk, we expect that in most cases LYS and GLU will retain their hydration shells.
In accord with this observation, it is found that LYS and GLU in
A$\beta$-peptides are solvated in bulk water \cite{Klimov_Thirumalai_Struc03,Tarus_Thirumalai_JACS06}.
 The orientation dependent interactions in the nanopore show that confinement
(Fig.~\ref{fig:Lys-Glu}b) can promote salt bridge formation between LYS and GLU.
Fig.~\ref{fig:Lys-Glu}c shows that the LYS-GLU system is largely insensitive to
the solvent density.
Even with the net solvent density reduced by 50\%, both the
$\rightarrow \leftarrow$ and the $\uparrow \uparrow$ interactions are similar
to those at bulk solvent density (Fig.~\ref{fig:Lys-Glu}b).
The distant minimum at $\sim 2.3$ nm for the $\rightarrow \leftarrow$
orientation disappears with the decrease in solvent density.

\textbf{Confinement enhances peptide stability:}
Our primary motivation for undertaking these studies was to gain insights into the stability
of proteins in confined spaces.
The extent of confinement-induced stabilization (or destabilization) of
peptides or proteins, with respect to bulk, will depend on the peptide sequence,
entropy loss of the denatured state ensemble (DSE) and solvent density.
(1) While nonpolar SCs prefer to be localized near the hydrophobic pore
boundary, oppositely charged side chains can be stabilized in contact either by
the formation of a salt bridge or by interior solvation.
Confinement also drives polar SCs to form contacts.
(2) Confinement in a narrow cylinder, as is the case in ribosomes, also
restricts the rotamer degrees of freedom, thus selecting a particular orientation
that may not be preferred in the bulk.
The decrease in the rotamer degrees of freedom in nanopores, as well as overall
restrictions in the conformations of the peptide increases the entropy of the DSE.
(3) Confinement destabilizes the contact minimum between hydrophobic SCs only if
the water density in the nanopore and the bulk are similar.
At lower densities the contact minimum is favored \cite{Vaitheeswaran_PNAS08}.
These observations suggest that for a generic polypeptide sequence, confinement
should enhance the stability of ordered states.

To test these predictions, we simulated two tripeptides, (Ala)$_3$, a
hydrophobic sequence and Lys-Ala-Glu, a hydrophilic sequence with no net charge,
in bulk water and in two cylindrical pores (Fig.~\ref{fig:peptides_snap}).
Both peptides are in their zwitterionic form, terminated with a
positively charged amino group and a negatively charged carboxylic group.
In the pores, we find that the center of mass of the (Ala)$_3$ backbone is
close to the surface.
The charged backbone is hydrated, while the nonpolar side chains are sequestered
from the water.
Not surprisingly, the center of mass of the hydrophilic sequence remains close
to the pore axis, away from the surface.
These findings are in accord with our expectations based on interactions
between amino acid side chains in cylindrical nanopores.

The changes in stability of the peptide conformations, with respect to the bulk,
can be assessed by computing the distributions of the backbone dihedral angles
$\phi$ and $\psi$ of the middle residue in each peptide
(Fig.~\ref{fig:rama_contours}).
The $\alpha$-helical region is defined as $-80^\circ \le \phi \le -48^\circ$ and
$-59^\circ \le \psi \le -27^\circ$ \cite{Klimov_Thirumalai_Struc03}.
We find that cylindrical confinement stabilizes both peptides, with the
hydrophilic sequence being more strongly stabilized (See Table~\ref{tab:helix}).
The extent of stability clearly depends on the sequence and its length, pore
geometry and solvent density, just as predicted using the results for PMFs
between amino acid side chains.

\section{Conclusions}

Model systems, such as hydrocarbon chains that can adopt compact structures in water, provide useful caricatures of the earliest stages of collapse transition in proteins. It is now firmly established that only for hydrocarbon chains longer than $n \sim$ 20 the collapsed structure is stable. The preference for the bonds to be in all trans configuration prevents smaller hydrocarbon chains (at least sizes on the order of dodecane) from adopting globular conformation. Surprisingly, the collapsed structure is ordered. The chains adopt  \textbf{HH} and higher order structures such as toroids depending on the length of the hydrocarbon chains.  Similarly, association between two hydrocarbon chains produces an ordered structures in which each chain adopts a 1.5 turn \textbf{HH}. It appears that in both processes water expulsion occurs early (before the chains adopt regular structures) although much more exhaustive simulations are required to quantify the time scales and the mechanisms of assembly.

Ion solvation in confined space not only reveals an asymmetry between positive and negative but also shows the importance of charge density, $\zeta$, in the preference for interior and surface solvation.  Charge density, a single parameter, accounts for excluded volume and electrostatic effects. Our work predicts that lower charge density ions prefer the surface and as the charge density increases interior solvation occurs. For a given $\zeta$ anions have greater preference for the surface than cations.  Most studies (see for example \cite{Jungwirth_Tobias_JPCb02}) have argued that ion polarization has to be accounted for in order to explain the enhanced propensity of I$^{-1}$ to be at the air/water interface than Cl$^{-1}$. Although polarization and size are not unrelated our work qualitatively explains in terms of $\zeta$ alone the tendencies of ions of differing sizes to localize near the surface.

An important aspect of the current perspective is to provide molecular basis for explaining stability of encapsulated peptides in cylinders. First, side chain orientation can dramatically alter interactions in confinement compared to bulk. Second, interactions between charged residues (lysine and glutamic acid for example) in confined space depends strongly on a balance between hydrophobic and ionic interactions. Using these results and loss of greater conformational entropy of the unfolded states relative to the folded structures upon confinement we have shown that for a generic amphiphilic sequence folded states should be stabilized when trapped in cylindrical pores. Explicit simulations in water  \cite{Vaitheeswaran_PNAS08} and extensive coarse-grained simulations of peptides in carbon nanotubes \cite{OBrien08NanoLett} support the theoretical expectation.

 \begin{figure}[htbp] 
   \centerline{\includegraphics[width=0.8\textwidth]{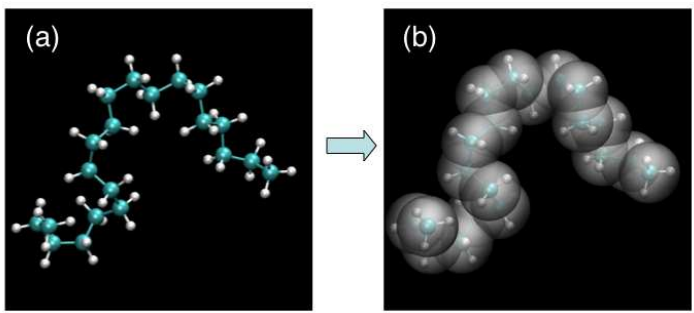}}
   \caption{(a) Stick representation of a HC chain ($CH_3-(CH2)_{18}-CH_3$) (b) Each methane group is
represented with a bead centered at the carbon atom.}
   \label{fig:model}
 \end{figure}

 \begin{figure}[htbp]
   \centerline{\includegraphics[width=0.8\textwidth]{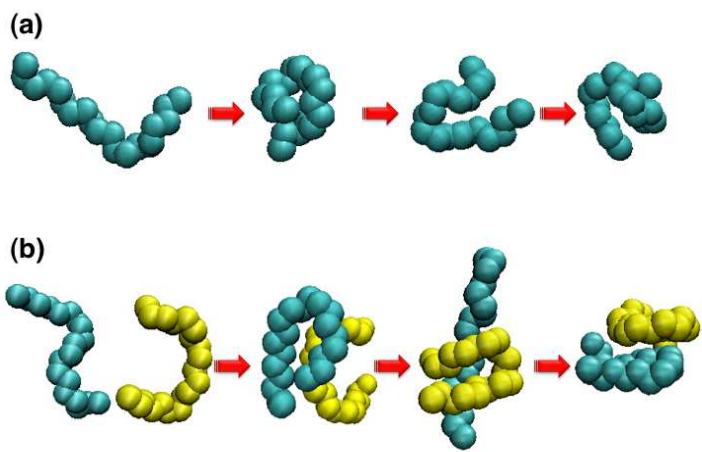}}
   \caption{(a) Representative snapshots of a trajectory showing a single HC chain collapses in water (b) Snapshots of a trajectory showing two HC chains collapse and associate in water. In both cases the HC chains adopt ordered helical hairpin structures.}
   \label{fig:snapshots}
 \end{figure}

 \begin{figure}[!htbp]
   \centerline{\includegraphics[width=0.6\textwidth]{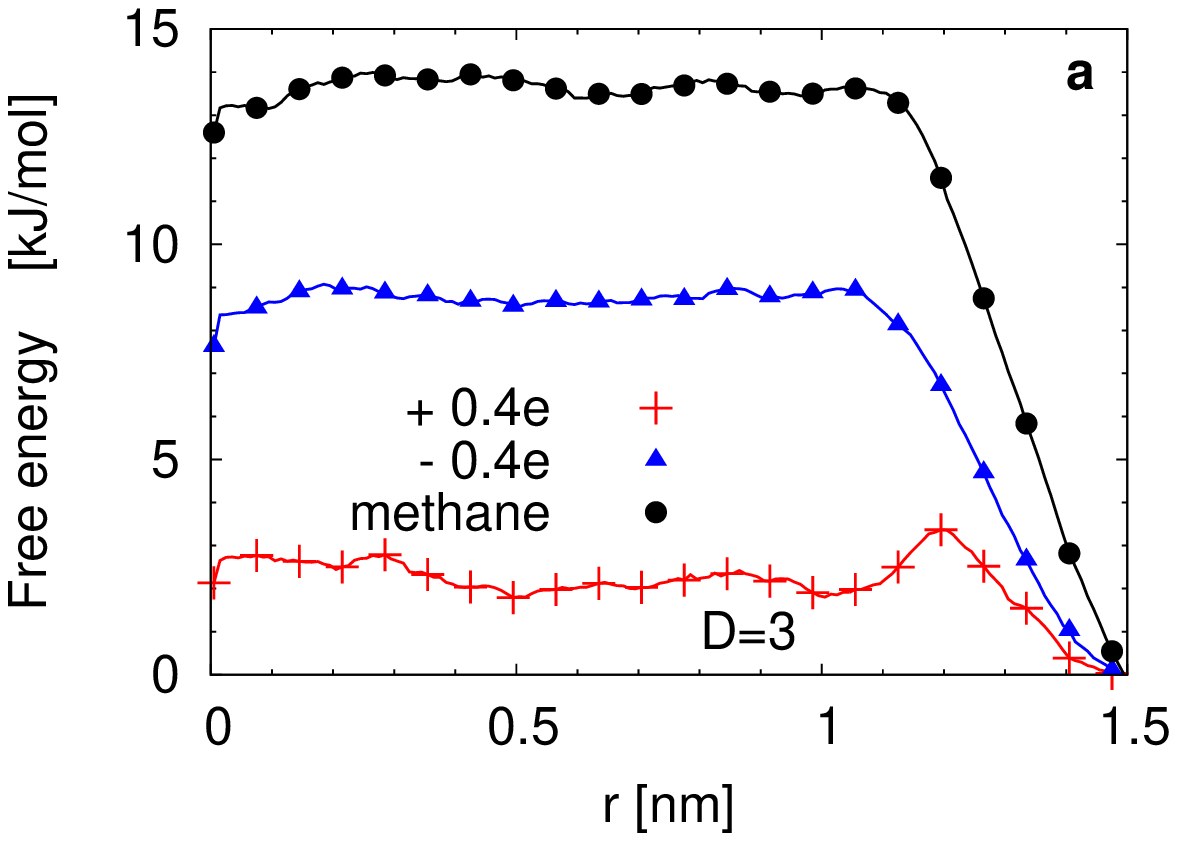}}
   \centerline{\includegraphics[width=0.6\textwidth]{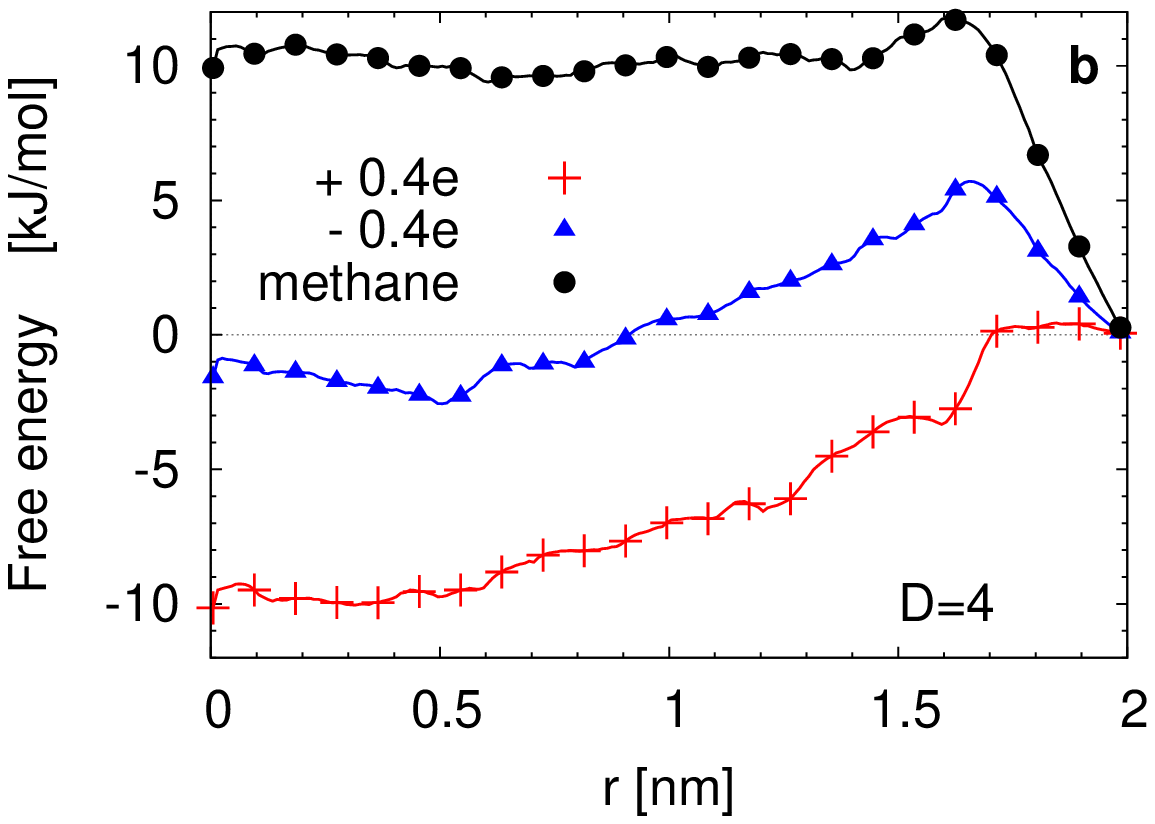}}
   \centerline{\includegraphics[width=0.6\textwidth]{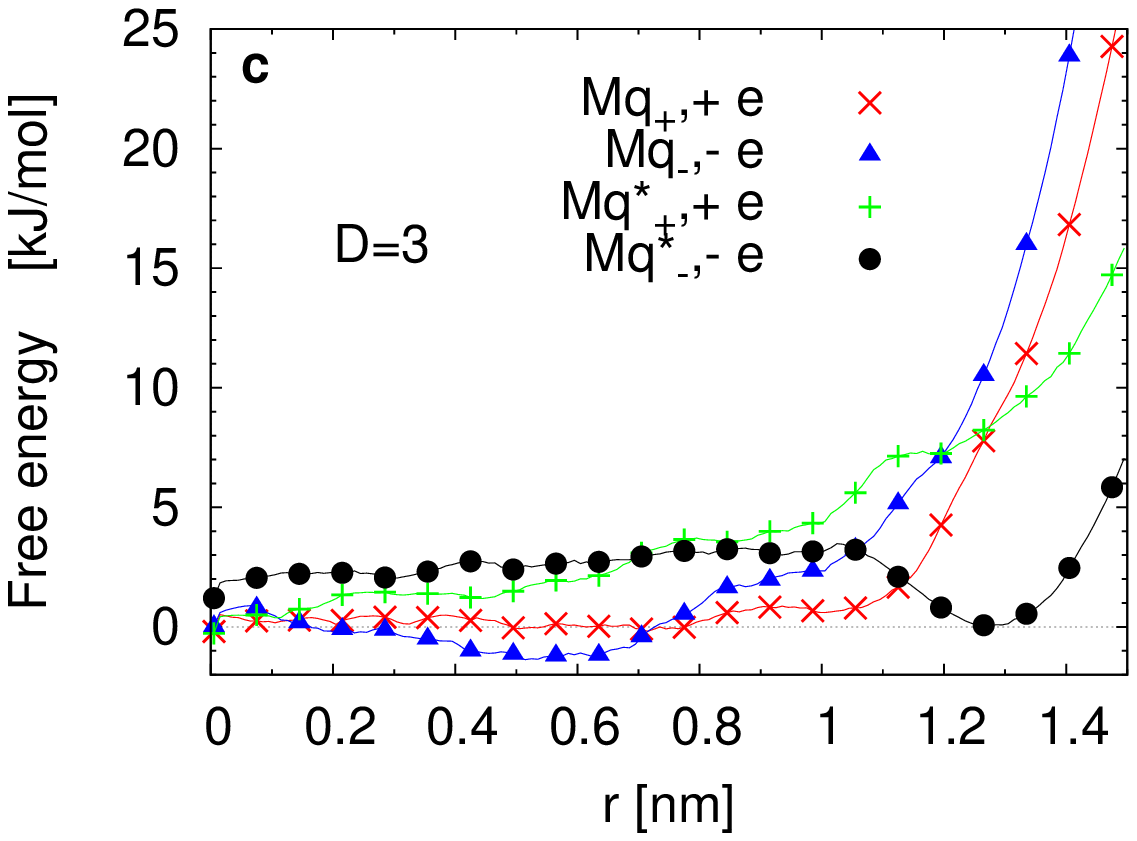}}
   \caption{ (a,b) Free energies of methane molecules and ions M$_{q\pm}$, derived by
 assigning charges of $\pm 0.4e$ to the methanes, in spherical water droplets
 of diameter $D$ nm.
 The zero of the free energy scale for every droplet is at the surface.
 The methanes have a strong preference for the droplet surface at all values
 of $D$.
 In the case of the ions, the surface propensity depends strongly on droplet size.
 In every droplet, anions have a greater preference for the surface than cations
 of the same charge magnitude.
 (c) Free energies profiles for ions M$_{q+}$, M$_{q-}$, M*$_{q+}$ and
 M*$_{q-}$ of charge magnitude $1.0e$. The starred ions have twice the volume and
 therefore half the charge density of the corresponding unstarred ion.
 Curves for M$_{q+}$, M$_{q-}$ and M*$_{q+}$ are referenced to the origin 
 (droplet center) while that for M*$_{q-}$ is referenced to the minimum at
 1.3 nm for clarity.
 In all cases, only differences within each curve are relevant.}
   \label{fig:free_energy_D1-3}
 \end{figure}

 \begin{figure}[htbp]
   \centerline{\includegraphics[width=0.6\textwidth]{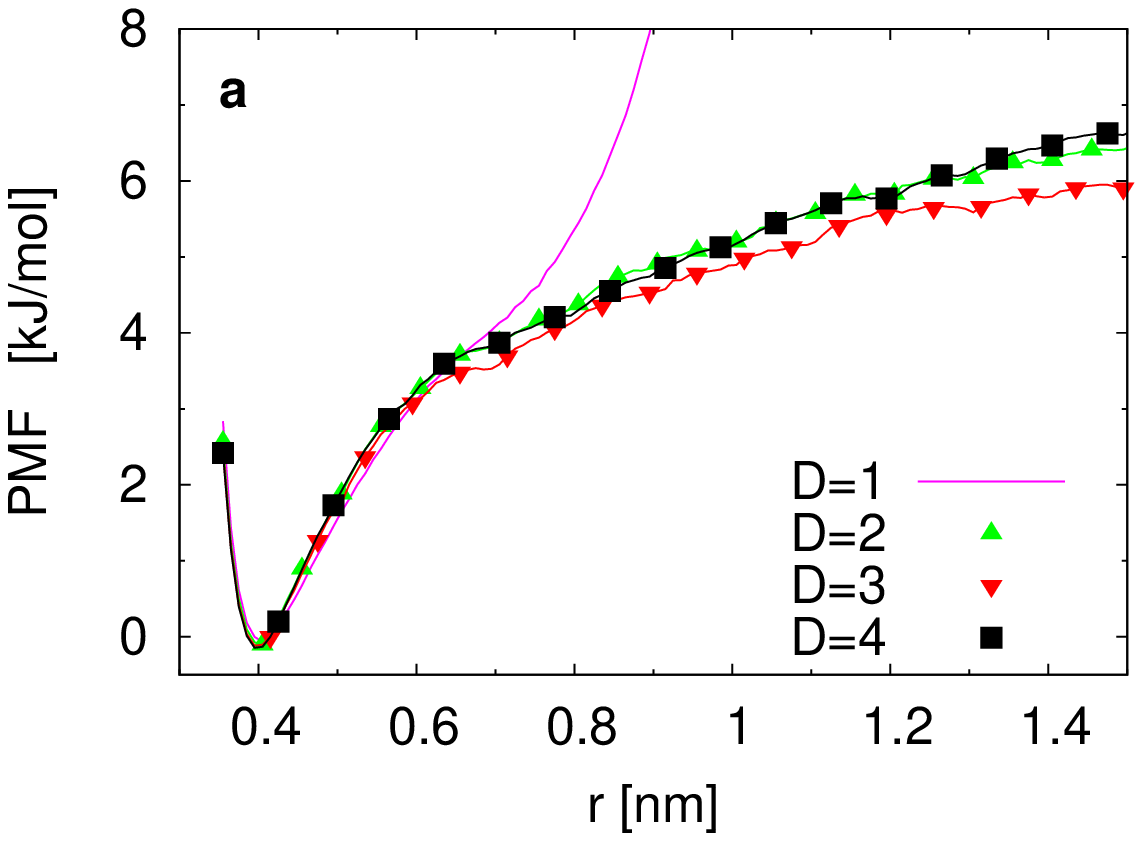}}
   \centerline{\includegraphics[width=0.6\textwidth]{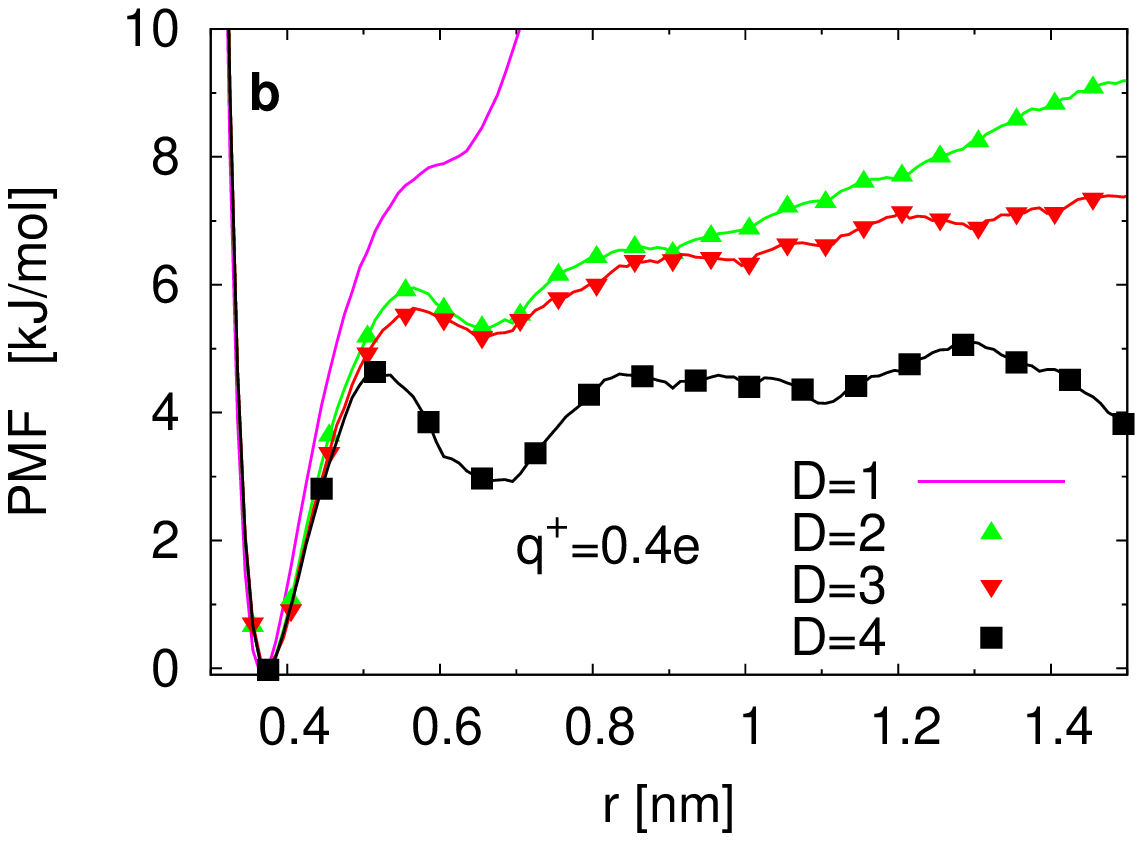}}
   \caption{(a) Potentials of mean force (PMFs) between two methane molecules in spherical droplets of
 diameter $D$ nm.
 Just as in the bulk, there is a distinct primary minimum.
 However, the characteristic solvent separated minimum is absent even at
 $D=4$ nm.
 The curves are shifted vertically so that the zero of the free energy scale
 is at contact for the two methanes.
 (b) PMFs between M$_{q+}$ and M$_{q-}$, with a charge magnitude of $0.4e$, in
 droplets of different sizes.}
   \label{fig:pmf}
 \end{figure}

 \begin{figure}[htbp]
   \centerline{\includegraphics[width=0.6\textwidth]{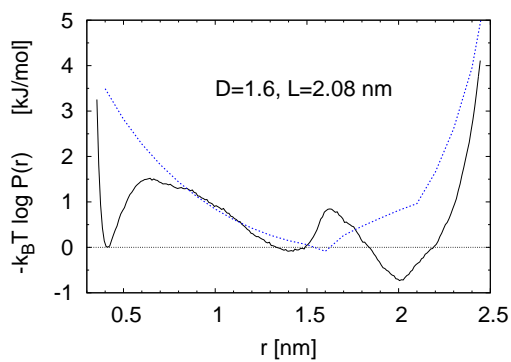}}
   \caption{Interaction free energies of two methane molecules in a cylindrical nanopore
 of diameter $D$ and length $L$ nm, containing water at bulk density.
 The dotted blue line shows the contribution from the translational entropy of
 the hydrophobic species, assuming that they are strictly confined to the pore surface.}
   \label{fig:cyl_d1.6}
 \end{figure}

 \begin{figure}[htbp]
   \centerline{\includegraphics[width=0.6\textwidth]{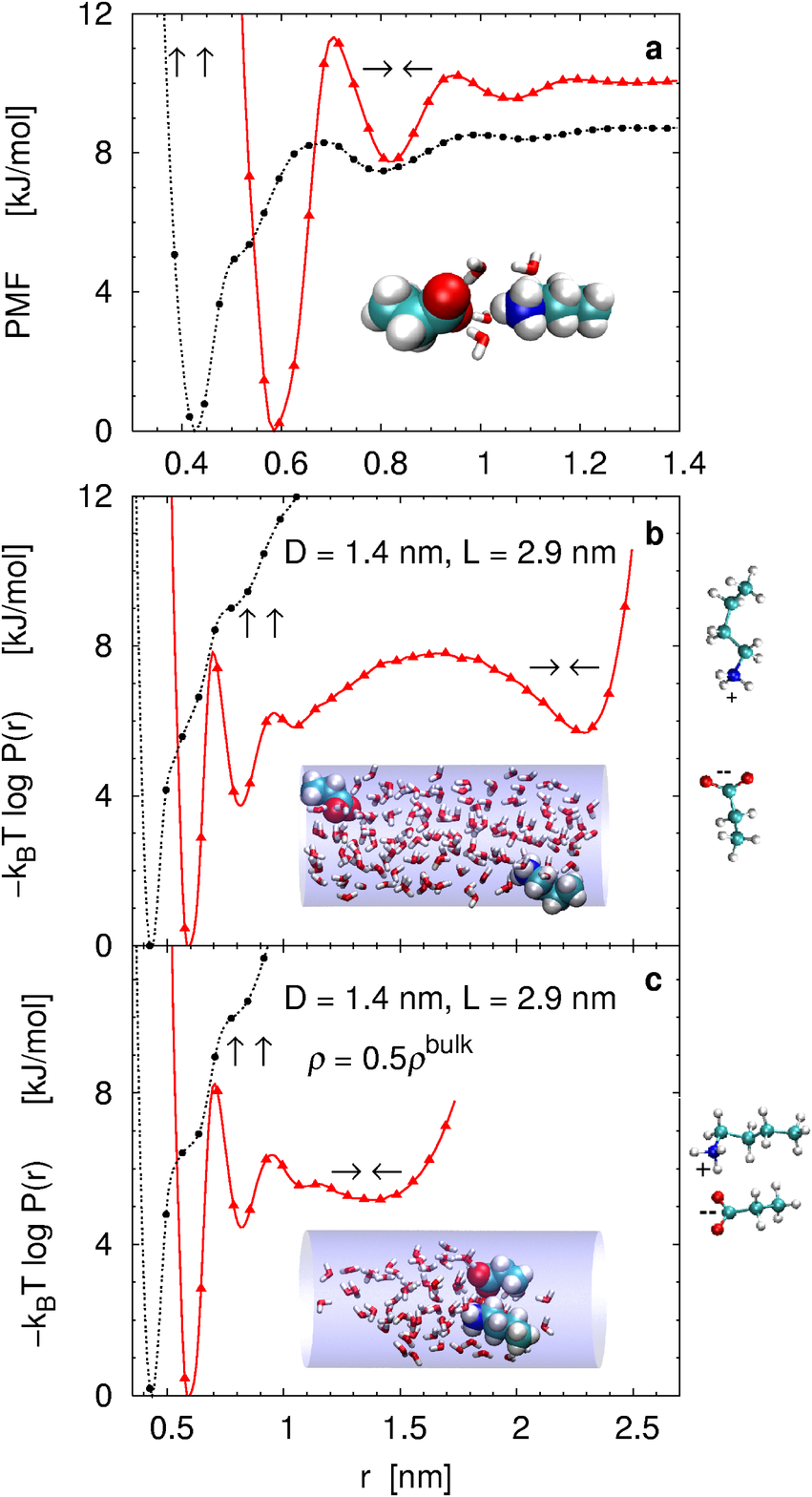}}
   \caption{(a) Potentials of mean force between LYS and GLU in bulk water in the
 $\rightarrow \leftarrow$ (stick figure in b), and $\uparrow \uparrow$
 (stick figure in c) orientations.
 Curves are translated vertically so that the zero of the free energy scale
 is at contact.
 The snapshot shows the $\rightarrow \leftarrow$ pair at the solvent separated
 minimum.
 (b) Free energies of interaction of LYS and GLU in a nanopore of diameter $D$
 and length $L$. Water is at bulk density.
 The inset shows the $\rightarrow \leftarrow$ pair at the distant minimum at
 $\sim 2.3$ nm.
 (c) LYS-GLU interaction free energies in the same nanopore at half the bulk
 water density. The inset shows the $\uparrow \uparrow$ pair at contact.}
   \label{fig:Lys-Glu}
 \end{figure}

 \begin{figure}[htbp]
   \centerline{\includegraphics[width=0.6\textwidth]{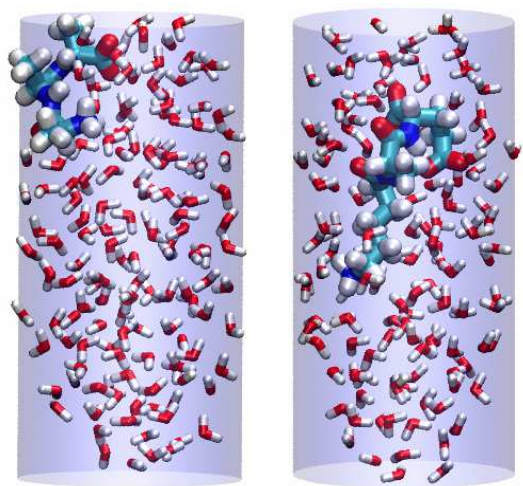}}
   \caption{The tripeptides (Ala)$_3$, a hydrophobic sequence (left) and Lys-Ala-Glu,
 a hydrophilic sequence with no net charge (right), in cylindrical pores with
 water at bulk density.}
   \label{fig:peptides_snap}
 \end{figure}

 \begin{figure}[htbp]
   \centerline{\includegraphics[width=5in]{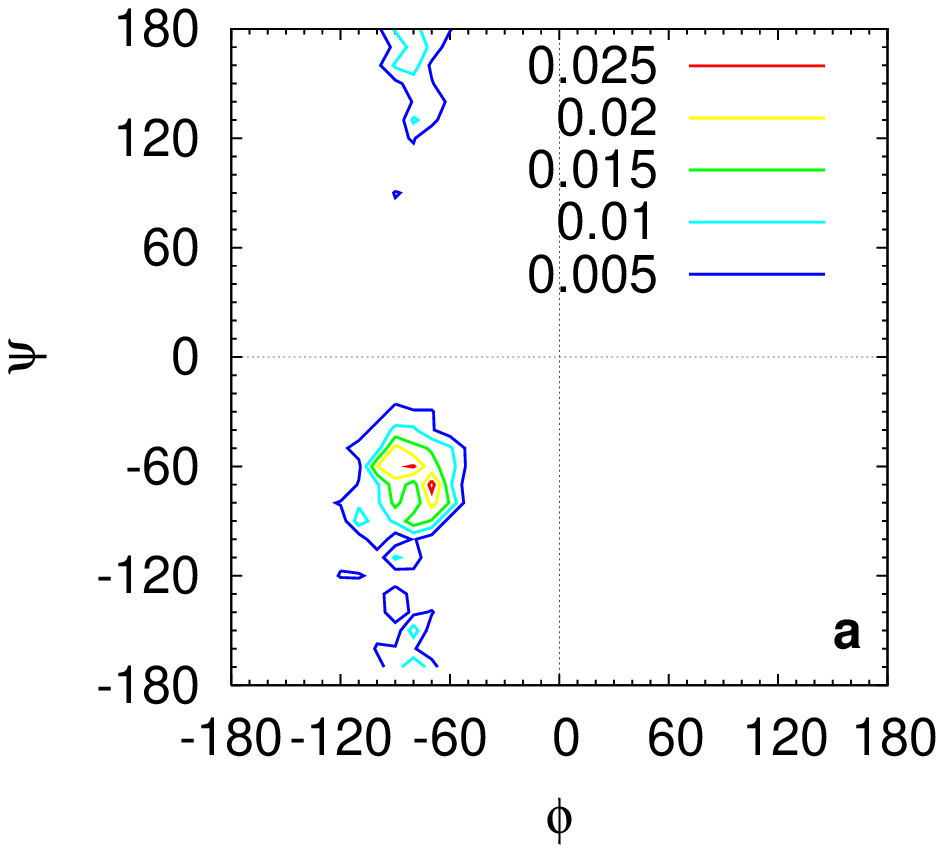}}
   \centerline{\includegraphics[width=5in]{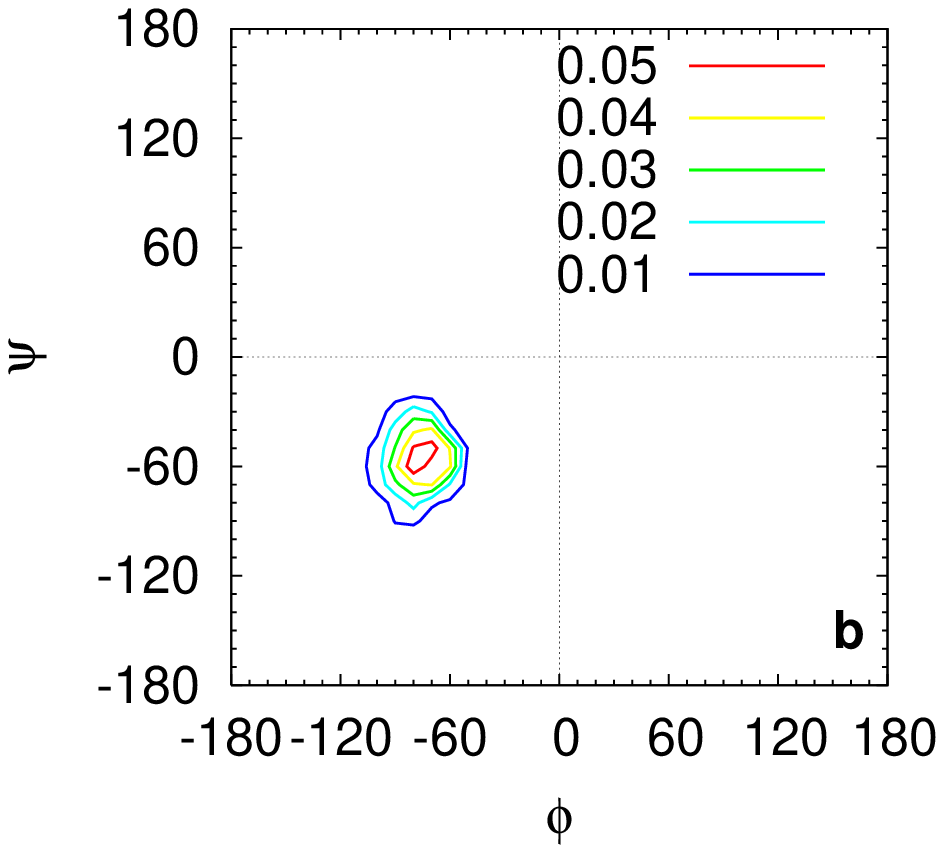}}
   \caption{Probability contours in $\phi-\psi$ space of the backbone, for the middle
 residue of Lys-Ala-Glu, in (a) bulk water ($P^B$) and (b) the nanopore ($P$)
 with interior water at bulk density.
 The changes in the stability of the ordered states are computed using
 $P_\alpha$ ($P_\alpha^B$), where $P_\alpha$ ($P_\alpha^B$) is the number of
 points in the $\alpha-$helical region defined by
 $-80^\circ \le \phi \le -48^\circ$ and $-59^\circ \le \psi \le -27^\circ$.}
   \label{fig:rama_contours}
 \end{figure}

\begin{table}[!htbp]
 \caption{Stability changes in peptides confined to cylindrical pores}
 \label{tab:helix}
 \setlength{\tabcolsep}{5mm}%half the column spacing
\begin{minipage}{\textwidth}
 \begin{tabular}{@{\vrule height 10.5pt depth4pt  width0pt}ccc}
  \hline\hline
  Pore size $(D,L)$\footnote{$D$ and $L$ are the diameter and length of cylindrical pores in nm.} &
  $\Delta F$ [k$_B$T]\footnote{$\Delta F = -\log(P_{\alpha} / P^{\mathrm{B}}_{\alpha})$ for (Ala)$_3$.
                                $P_{\alpha}$ and $P^{\mathrm{B}}_{\alpha}$ are the probabilities of being
                                in the $\alpha$-helical region in the pore and bulk respectively.} &
  $\Delta F$ [k$_B$T]\footnote{$\Delta F$ for Lys-Ala-Glu.}\\\hline
  2.0, 2.9            & -0.61                   & -0.84 \\
  1.4, 2.9            & -0.14 (-0.19\footnote{$\Delta F$ for $\rho = 0.5 \rho^{\mathrm{bulk}}$}) & -1.50 (-1.02\footnotemark[4]) \\
  \hline
 \end{tabular}
\end{minipage}
\end{table}

\begin{acknowledgements}
We are grateful to Govardhan Reddy for useful discussions. This work is supported by the National Science Foundation (NSF CHE 09-14033).
\end{acknowledgements}

\bibliographystyle{spmpsci}      % mathematics and physical sciences
\bibliography{water_ref}   % name your BibTeX data base

%% Non-BibTeX users please use
%\begin{thebibliography}{}
%%
%% and use \bibitem to create references. Consult the Instructions
%% for authors for reference list style.
%%\end{thebibliography}

\end{document}